\documentclass[reprint,superscriptaddress,a4paper,aps,longbibliography,twoside]{revtex4-2}
\usepackage{graphicx}\graphicspath{{./},{/home/user/fig/publi/muon/}}
\usepackage{amsfonts}
\usepackage{revsymb}
\usepackage{amsmath}
\usepackage{amssymb}
\usepackage{bm}
\usepackage{xcolor}
\DeclareMathAlphabet\mathbfcal{OMS}{cmsy}{b}{n}
\begin{document}
\title{Experimental determination of the spin Hamiltonian of the  cubic chiral magnet MnSi}

\author{P. Dalmas de R\'eotier}
\affiliation{Universit\'e Grenoble Alpes, CEA, Grenoble INP, IRIG-PHELIQS, F-38000 Grenoble, France}
\author{A. Yaouanc}
\affiliation{Universit\'e Grenoble Alpes, CEA, Grenoble INP, IRIG-PHELIQS, F-38000 Grenoble, France}
\author{G. Lapertot}
\affiliation{Universit\'e Grenoble Alpes, CEA, Grenoble INP, IRIG-PHELIQS, F-38000 Grenoble, France}
\author{C. Wang}
\affiliation{Laboratory for Muon-Spin Spectroscopy, Paul Scherrer Institute,
CH-5232 Villigen-PSI, Switzerland}
\author{A. Amato}
\affiliation{Laboratory for Muon-Spin Spectroscopy, Paul Scherrer Institute,
CH-5232 Villigen-PSI, Switzerland}
\author{D. Andreica}
\affiliation{Faculty of Physics, Babes-Bolyai University, 400084 Cluj-Napoca, Romania}

\date{\today}

\begin{abstract}
  
  A thorough description of the physics of a magnetic compound requires the validation of its microscopic spin Hamiltonian. Here, from the analysis of muon-spin rotation spectra recorded in the magnetically ordered state at low temperature in zero and finite magnetic fields, we determine the minimal  Hamiltonian for the chiral binary intermetallic magnet MnSi, consistent with its high-temperature nonsymmorphic cubic space group  P$2_1$3. The model provides constraints for the orientation of the Moriya vector characterizing the microscopic Dzyaloshinskii-Moriya interaction, with respect to the Mn nearest-neighbor bonds. Small twist and canting of the magnetic structure are revealed. Our result indicates that, within experimental uncertainties, the magnetoelastic coupling is not strong enough to lower the paramagnetic crystal symmetry in the magnetically ordered state. Additional implications from our work are discussed and complementary studies are suggested.
\end{abstract}

\maketitle

{\it Introduction}.
Helices appear in different systems of condensed matter and biology \cite{Kornyshev07}. Sometimes rather complex patterns are formed like in the blue phases of liquid crystals \cite{Wright89} and the skyrmion textures in chiral magnets with B20 crystal structure \cite{Cheong22} as first discovered for MnSi \cite{Muhlbauer09a}.  The similarity in the physics of those systems is striking \cite{Tewari06,Hamann11}.  

The binary intermetallic compound MnSi is an exciting playground that is boosted by the availability of large and high purity single crystals --- see, for example, Ref.~\onlinecite{Doiron03} --- and its properties can be studied in convenient temperature, magnetic field and pressure ranges \cite{Andreica10}. While still exotic with its lack of inversion symmetry, MnSi is a relatively simple cubic compound (nonsymmorphic space group  P$2_1$3) with four symmetry-equivalent manganese atoms in the unit cell. This explains the appreciable number of theoretical works published since the 1980s which have been based on a continuum description assuming a strong ferromagnetic interaction  with an additional weak chiral term described by a scalar parameter \cite{Bak80,Nakanishi80,Belitz06,Maleyev06}.  Microscopic models have only been considered more recently \cite{Hopkinson09,Hamann11,Chizhikov12,Borisov21,Hall21} in a more limited number of studies. 

The compound has a long experimental history beginning with the determination of its crystal structure at room temperature in 1933 \cite{Boren33}. Magnetic measurements indicate a magnetic phase transition at temperature $T_{\rm c} \simeq  30$~K~\cite{Williams66} with a weak first order character \cite{Date77,Stishov07}. The spin structure is helicoidal according to nuclear magnetic resonance (NMR) \cite{Motoya76}. 

Experimental and theoretical works \cite{Ziebeck82,Yaouanc20,Chen20,Choi19,Dalmas21,Fang22} suggests MnSi to be a dual system with itinerant and localized electronic subsets. This dual picture seems to be widespread, since it applies, for example, to the ferromagnetic superconductor UGe$_2$ \cite{Yaouanc02,Sakarya10,Haslbeck19}, the strange metal regime of cuprate superconductors \cite{Ayres21}, or a Ce heavy fermion system \cite{Machida22}. In the weak itinerant ferromagnet MnSi, the duality is thought in terms of a Hund metal, in which interorbital exchange interactions (Hund's coupling) give rise to strong ferromagnetic correlations betwen the electronic subsets \cite{Georges13,Nomura22}. This Hund metal character could also apply to the sibling compound FeGe and explain the failure of a single subset viewpoint for this system \cite{Grytsiuk19}. 

The propagation wavevector ${\bf k}$ of the magnetic structure is so small that it is most easily measured with small angle neutron scattering (SANS) \cite{Ishikawa76}. This technique evidences equivalent magnetic satellites only in the vicinity of the reciprocal space origin. Hence, no conventional magnetic  structure refinement can be achieved \cite{Rossat87}. Since the original NMR and SANS  measurements in 1976, the Mn magnetic moments had been assumed to draw a helix around ${\bf k}$ in zero magnetic field and to form a conical phase of axis ${\bf k}$ under a modest field.

This simple picture was shown in 2016 to be a first approximation to the actual structure in zero field (ZF) at low temperature \cite{Dalmas16}, and later on, in the conical phase near $T_{\rm c}$ \cite{Dalmas17}. This result was derived from the analysis of spectra recorded  with the muon-spin-rotation ($\mu$SR) technique, within the framework of Bertaut's representation theory for magnetic structures \cite{Bertaut63,*Bertaut68,*Bertaut71,*Bertaut81}. In particular, a ZF double-helix structure was unveiled with one of the four Mn magnetic moments of the unit cell drawing an helix along ${\bf k}$ as one moves from cell to cell, while  the other three moments belong to a second helix that is appreciably twisted relative to the first one. 

While Bertaut's theory is nowadays a conventional tool for the determination of magnetic structures using  diffraction patterns --- see, for example, Refs.~\onlinecite{Wills01,Kenzelmann06,Kenzelmann08,Rodriguez19} for neutron data --- the number of remaining free parameters after its application  is still large.

Here, instead of the determination of parameters merely describing the magnetic structure, we consider a minimum nearest-neighbor spin Hamiltonian including the Heisenberg, Dzyaloshinskii-Moriya (DM), and Zeeman interactions. Assuming the possibility of small deviations from the regular helical or conical structure, a minimization of the energy is performed which imposes severe constraints on the actual magnetic structures. We fit the remaining free parameter to experimental $\mu$SR spectra recorded at 2~K in zero and 0.3~T fields oriented along the three principal directions of the cubic structure. This provides in turn quantitative information on the parameters entering the spin Hamiltonian, in particular, the Cartesian components of the microscopic Moriya vector.

{\it Some basic physical properties of MnSi}.
The lattice parameter is $a_{\rm latt}$ = 4.558~\AA. The Mn atoms occupy the $4a$ Wyckoff position which is entirely defined by parameter $x_{\rm Mn}$ = 0.138. As done previously  \cite{Dalmas16}, we shall specify the position of a unit cell by the cubic lattice vector $ {\bf i }$ and that of a Mn atom within a cell by $ {\bf d}_\gamma$ with $\gamma \in \{{\rm I},{\rm II},{\rm III},{\rm IV} \}$. For convenience, we list the ${\bf d}_\gamma$ coordinates in Table~\ref{table_coordinates}.

\begin{table}
  \caption{Coordinates of the Mn atoms in MnSi, corresponding to the equivalent sites of Wykoff position $4a$ in space group P2$_1$3. All the coordinates are expressed in unit of the lattice parameter.}
\label{table_coordinates}
\begin{tabular}{l|c}
$\gamma$ & ${\bf d}_\gamma$ \cr\hline
{\rm I} & $(x_{\rm Mn},x_{\rm Mn},x_{\rm Mn})$\cr
{\rm II} & $(\bar{x}_{\rm Mn}+\frac{1}{2}, \bar{x}_{\rm Mn}, x_{\rm Mn}+\frac{1}{2})$\cr
{\rm III} & $(\bar{x}_{\rm Mn}, x_{\rm Mn}+\frac{1}{2}, \bar{x}_{\rm Mn}+\frac{1}{2})$\cr
{\rm IV} & $(x_{\rm Mn}+\frac{1}{2}, \bar{x}_{\rm Mn}+\frac{1}{2}, \bar{x}_{\rm Mn})$\cr
\end{tabular}
\end{table}

From SANS it has been established that ${\bf k}$ in ZF is collinear to one of the four three-fold axes with an incommensurate modulus $k  \approx 0.345$~nm$^{-1}$ at low temperature \cite{Ishikawa76,Ishida85,Fak05,Grigoriev06,Muhlbauer09a} . In the conical phase, ${\bf k}$ is parallel to the external magnetic field ${\bf B}_{\rm ext}$ with approximately the same modulus.

{\it Spin Hamiltonian and its treatment}.
We consider classical spins ${\bf S}$ interacting through the ferromagnetic Heisenberg and DM interactions. The Hamiltonian writes
\begin{eqnarray}
{\mathcal H}  & = &   - \frac {1}{2}
  \sum_{\langle{\bf i},{\bf i}^\prime,\gamma,{\gamma^\prime}\rangle}
  J {\bf S}_{{\bf i}, \gamma}\cdot {\bf S}_{{\bf i}^\prime, \gamma^\prime} \label{Hamiltonian} \\
  & + & \frac {1}{2} \sum_{\langle{\bf i},{\bf i}^\prime, \gamma,{\gamma^\prime}\rangle}
   {\bf D}_{{\bf i},\gamma; {\bf i}^\prime, \gamma^\prime} \cdot
    \left ( {\bf S}_{{\bf i}, \gamma}\times {\bf S}_{{\bf i}^\prime,\gamma^\prime} \right )  +  \sum_{{\bf i},\gamma} g\mu_{\rm B} {\bf S}_{{\bf i},\gamma} \cdot {\bf B},\nonumber
\end{eqnarray}
where the first two sums are limited to nearest-neighbors \cite{SM}. The last quantity is the Zeeman term, in which the magnetic induction ${\bf B}$ is related to ${\bf B}_{\rm ext}$ through the demagnetization field. The spectroscopic factor is set to the experimental value $g \approx 2$ \cite{Date77,Demishev11}. In Eq.~\ref{Hamiltonian}, ${\bf D}_{{\bf i},\gamma;{\bf i}^\prime,\gamma^\prime}$ represents the Moriya pseudovector (or axial vector) associated with atomic bond between sites ${\bf i},\gamma$ and ${\bf i}^\prime,{\gamma^\prime}$. It is invariant through cubic lattice translations. With four Mn spins in the unit cell and six neighbor spins for each of them, we have twenty-four different ${\bf D}_{{\bf i},\gamma;{\bf  i}^\prime,\gamma^\prime}$ vectors. They are related to each other by the antisymmetry relation ${\bf D}_{{\bf i}^\prime, \gamma^\prime;{\bf i},\gamma} = -{\bf D}_{{\bf i},\gamma;{\bf i}^\prime,\gamma^\prime}$ and the symmetry elements of point group 23 \cite{SM}. In fact the specification of a single ${\bf D}_{{\bf i},\gamma;{\bf i}^\prime,\gamma^\prime}$ vector suffices to determine the whole set. We have chosen bond I--II as the reference bond such that ${\bf D}_{{\bf i},{\rm I};{\bf i},{\rm II}} \equiv {\bf D} =(D^x,D^y,D^z)$ where the Cartesian components are expressed in the cubic reference frame. Note that the DM Hamiltonian can equivalently be written as the weighted sum of three antisymmetric invariants \cite{Hall21}, the weighting factors being the ${\bf D}$ components.  

To lower the energy, the magnetic structure is allowed to slightly deviate from the regular helical or conical structure through twist and canting angles. We will explicit these angles thereafter. Assuming the product $k a_{\rm latt}$, the twist and canting angles, and the $D^\alpha/J$ ratios to be small parameters, the energy is written as an expansion up to second order in these quantities. Thanks to the incommensurate nature of the magnetic ordering, the energy minimization can be performed analytically \cite{SM,Chizhikov12}. It leads to
\begin{eqnarray} 
\frac{2}{3}\,\frac{-D^x + D^y -2 D^z}{J} & = & -k a_{\rm latt},
\label{MnSi_micro_k}
\end{eqnarray}
and to analytical expressions for the twist and canting angles which depend on the Mn sublattice, the orientation of ${\bf k}$ and parameter
\begin{eqnarray}
\frac{D^x + D^y}{J}  & \equiv & \sigma.
\label{MnSi_micro_varepsilon}
\end{eqnarray}
The resulting magnetic structure is found consistent with the prescriptions of representation analysis, as expected. Note that Eq.~\ref{MnSi_micro_k} replaces the continuous-field model relation $k = |{\mathcal D}|/B_1$ linking $k$ with the scalar ${\mathcal D}$ describing the DM interaction and the exchange stiffness $B_1$; see, e.g., Ref.~\onlinecite{Bak80}.

{\it The polarization function}.
A $\mu$SR experiment gives access to a polarization function, i.e.\ the time evolution $P_{Z,X}(t)$ of the projection of the muon spins along the direction of the beam polarization ($Z$) or a direction perpendicular to ${\bf B}_{\rm ext}$ ($X$) \cite{Yaouanc11}. As a first step towards its computation, we need an expression for Mn magnetic moment ${\bf m}_{{\bf i},\gamma}$ = $-g\mu_{\rm B} {\bf S}_{{\bf i},\gamma}$ at position ${\bf r}_{{\bf i},\gamma}$. Consistently with the helical or conical magnetic structure \cite{Dalmas16,Dalmas17}:
\begin{eqnarray}
{\bf m}_{\ell;{\bf i},\gamma} & = & m_\perp \left[  \cos \left({\bf k}_\ell\cdot {\bf r}_{{\bf i}, \gamma} \right) {\bf a}_{\ell,\gamma} 
 - \sin \left( {\bf k}_\ell\cdot {\bf r}_{{\bf i},\gamma} \right) {\bf b}_{\ell,\gamma} \right] +{\bf m}_\parallel,\cr & &
\label{moment_muon_general}
\end{eqnarray}
where ${\bf m}_\parallel $ is the projection of the magnetic moment along ${\bf B}_{\rm  ext}$ when a field is applied. The subscript $\ell$ labels one of the K-domains, and ${\bf a}_{\ell,\gamma}$ and ${\bf b}_{\ell,\gamma}$ are orthogonal unit vectors. In the regular helical and conical phases, vectors ${\bf a}_{\ell,\gamma}$ and ${\bf b}_{\ell,\gamma}$ are perpendicular to ${\bf k}_\ell$ and independent of $\gamma$. Here, we do not enforce these conditions. Instead, the minimization of the energy is obtained by allowing ${\bf a}_{\ell,\gamma}$ and ${\bf b}_{\ell,\gamma}$ to deviate from the ${\bf a}_\ell$ and ${\bf b}_\ell$ vectors of the regular structures \cite{SM}, which together with ${\bf k_\ell}/k$ form a direct orthonormal basis. The vectors ${\bf a}_{\ell,\gamma}$ and ${\bf b}_{\ell,\gamma}$ are deduced from ${\bf a}_\ell$ and ${\bf b}_\ell$ after two successive rotations \cite{Chizhikov12}. The first one, corresponding to a structure twist, is a rotation of angle $\omega_{\ell,\gamma}$ around ${\bf k}_\ell$. The second one, defining the structure canting, is a rotation around an axis ${\bm \Gamma}_{\ell,\gamma}$ perpendicular to ${\bf k}_\ell$.

Equipped with Eq.~\ref{moment_muon_general} and the prior determination of the muon crystallographic site and coupling parameter \cite{Amato14,Dalmas18}, the magnetic field vector ${\bf B}_{\rm loc}$ at the location of the probe can be derived for a given site in a given magnetic domain.
Then the evolution ${\bf S}_\mu(t)$ of the muon spin is computed from the Larmor equation $\frac{{\rm d}{\bf S}_\mu}{{\rm d}t} = \gamma_\mu {\bf S}_\mu\times {\bf B}_{\rm loc}$, where $\gamma_\mu$ = 851.6~Mrad\,s$^{-1}$\,T$^{-1}$ is the muon gyromagnetic ratio. Accounting for the spin-spin and spin-lattice relaxation rates, and averaging over the crystallographically equivalent muon sites in the crystal and over the magnetic domains, the model $P_{Z,X}(t)$ functions are computed \cite{SM}.

\begin{figure}
\begin{picture}(255,338)
\put(0,0){\includegraphics[width=\linewidth]{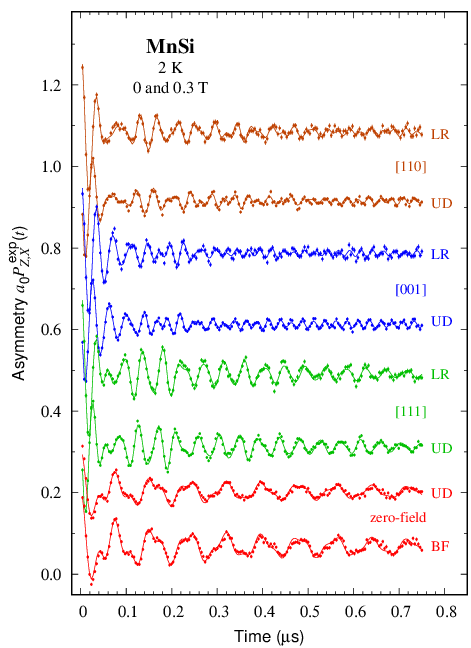}}
\put(125,278){
\includegraphics[scale=0.7]{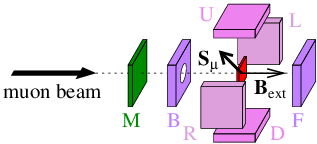}}
\end{picture}
\caption{$\mu$SR spectra of a MnSi crystal at 2~K, measured either in zero or in a 0.3~T field applied along the [111], [001], or [110] crystallographic direction as indicated in the figure. The ZF spectra obtained for the three orientations of the crystals are similar. For each measurement, the results corresponding to the relevant pair of detectors are displayed, with a vertical shift of 0.15 for better visibility. The full lines represent a combined fit to the different sets of data. The inset sketches the muon (M) and positron forward, backward, up, down, left and right (F, B, U, D, L, R) detectors with the sample in the middle of the spectrometer. ${\bf S}_\mu$ denotes the spin of the implanted muons.}
\label{spectra}
\end{figure}

\begin{figure}
\includegraphics[width=\linewidth]{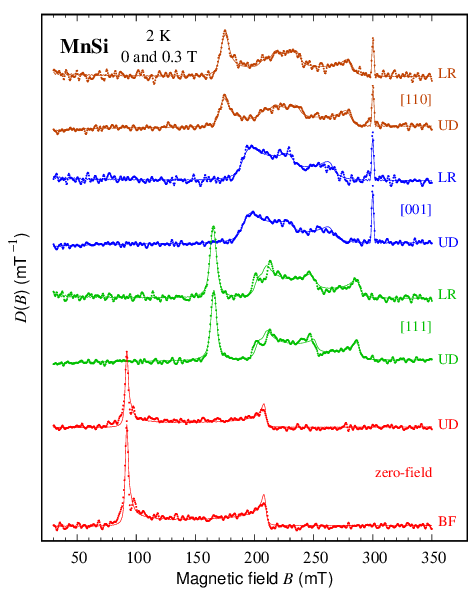}
\caption{Distribution of fields probed by the muons given by the Fourier transforms of the data and fits displayed in Fig.~\ref{spectra}. The narrow peaks visible at 300~mT for the [001] and [110] spectra correspond to a background of muons stopped out of the crystal.}
\label{FT}
\end{figure}

{\it Experiments and results}.
The $\mu$SR experiments were carried out with single crystals grown from Czochralski pulling; see Refs. \onlinecite{Dalmas18, Yaouanc20} for more details. The crystal were cut in the form of slabs oriented perpendicular to the [111], [001], or [110] cubic axes. The measurements were performed with the general purpose surface-muon (GPS) spectrometer of the Swiss muon source, Paul Scherrer Institute, Villigen, Switzerland \cite{Amato17}. 
 We display in Fig.~\ref{spectra} the spectra recorded at $2$~K in ZF and in a field of $0.3$~T applied along each of the three principal directions of a cubic structure, together with the results of a fit of our model to the data. For reference, Fig.~\ref{FT} presents the field distributions computed from the experimental data and fits. Overall, the description is rewarding; only near 0.2~T for the $[111]$ direction, some details of the experimental data are not captured by the model (Fig.~\ref{FT}).

This data analysis yields the Mn magnetic moment $m$ and the angle of the conical structure; see Table~\ref{moment_parameter}. As expected, the $m$ values are reasonably independent of the field direction and intensity, and are consistent with the literature; see, e.g., Refs.~\onlinecite{Williams66}, \onlinecite{Wernick72}, and \onlinecite{Yaouanc20}. The value of $\sigma$ (Eq.~\ref{MnSi_micro_varepsilon}) derived from the fit is $\sigma = 0.017\,(3)$. As a byproduct, we  compute the twist and canting angles in ZF to be  $0.83 \,(10)$ and $0.40\,(7)$ degrees, respectively \cite{SM}. While a twist angle of 0.83$^\circ$ seems very small, it is however non negligible compared to the average rotation angle 2.6$^\circ$ of the magnetic moments belonging to neighboring $\langle 111\rangle$ Mn planes. The canting is approximately twice as large as found for the celebrated La$_2$CuO$_4$ case \cite{Thio88}. 

\begin{table}
  \caption{Results of the combined fit to the data: Mn magnetic moment $m \approx (m_\perp^2+m_\parallel^2)^{1/2}$ and angle $\theta = \arctan(m_\perp/m_\parallel)$ characterizing the conical structure for an applied field of 0.3~T. }
  \label{moment_parameter}
\begin{tabular}{l|c|c}
Field direction & Moment $m$ ($\mu_{\rm B}$) & Cone angle $\theta$ (deg.) \cr \hline
zero-field & 0.3881\,(2) & 90\,(--) \cr
[111] & 0.388\,(2) & 66.7\,(3) \cr
[001] & 0.396\,(3) & 63.5\,(7)\cr
[110] & 0.388\,(2) & 66.6\,(5)\cr
\end{tabular}
\end{table}

\begin{figure}
\begin{picture}(245,140)
\put(0,0){\includegraphics[width=\linewidth]{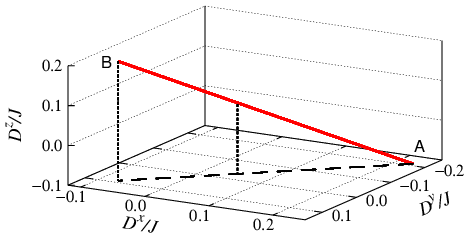}}
\put(145,73){\includegraphics[scale=0.375]{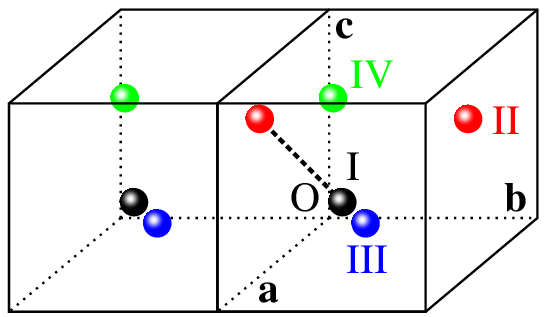}}
\end{picture}
\caption{Locus of the ${\bf D}$ vector Cartesian coordinates resulting from the data analysis. The components are normalized to the exchange parameter $J$. For the sake of clarity, the projection of the line onto the $D^z/J$ = $-0.1$ plane is shown as a dashed line. The plot is limited to ${\rm max}\left(|D^\alpha|/J\right)\lesssim 0.2$. The full line has been drawn for the nominal values of $k a_{\rm latt}$ and $\sigma$. Solutions in the vicinity of the line are possible according to the experimental uncertainties on each of these quantities. The vertical dotted line in the middle corresponds to the position of the minimum of $D/J$ or of the angle between ${\bf D}$ and the reference bond; see Fig.~\ref{DoJ_bond}. The inset depicts two unit cubic cells with the four types of Mn atoms and reference bond I-II.}
\label{DoJ_range}
\end{figure}

\begin{figure}
\includegraphics[width=\linewidth]{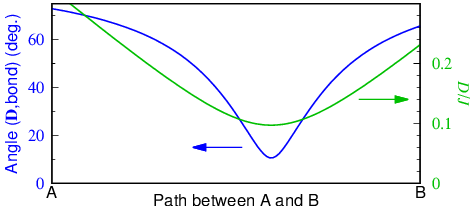}
\caption{Characterization of ${\bf D}$: (i) angle between the nearest-neighbor Mn atoms bond and the associated ${\bf D}$  and (ii) modulus of ${\bf D}$ normalized to $J$. Both the angle and $D/J$ are plotted for the path between the A and B points defined in Fig.~\ref{DoJ_range}. As for the preceding figure, solutions in the vicinity of the two lines are possible. The minimum angle is $\simeq 10.6$~degrees. 
}
\label{DoJ_bond}
\end{figure}

{\it Discussion and conclusions}.
We first discuss the quantitative information about ${\mathcal H}$ (Eq.~\ref{Hamiltonian}) deriving from our analysis. As explained above, from symmetry consideration, ${\mathcal H}$ only depends on four independent parameters: $J$, $D^x$, $D^y$ and $D^z$. A good estimate for $J$ is provided by the analysis of the temperature dependence of $m$ \cite{Yaouanc20,Dalmas21}: $J$ = 5.5\,(1)\,meV \footnote{In Refs.~\onlinecite{Yaouanc20} a continuous field model is used with a stiffness denoted as $B_1$. It is related to $J$ through $J$ = $4{\mathcal J}/3a_{\rm latt}^2$ \cite{Chizhikov12} with ${\mathcal J} \equiv B_1/2$. Numerically, from $B_1$  = $2.73\, (4) \times 10^{-40}$~J\,m$^2$, $J$ = $8.8 \,(1) \times 10^{-22}$~J.}.
Equations~\ref{MnSi_micro_k} and \ref{MnSi_micro_varepsilon} provide linear relations between $D^x$, $D^y$, and $D^z$. While a third relation linking them would be required for a complete determination, Fig.~\ref{DoJ_range} illustrates the set of ${\bf D}$ components consistent with the experimental data. The full line is obtained from the intersection of the planes defined by Eqs.~\ref{MnSi_micro_k} and \ref{MnSi_micro_varepsilon}. We note that the Moriya rules \cite{Moriya60} provide no information on the angle between ${\bf D}$ and the Mn nearest-neighbor bond, due to the absence of the requested symmetries. This is consistent with Fig.~\ref{DoJ_bond} for which no remarkable angle value appears. Figure~\ref{DoJ_bond} also displays the evolution of $D/J$ when ${\bf D}$ describes the full line of Fig.~\ref{DoJ_range}. Rewardingly, the condition $D/J \ll 1$ is fulfilled which is consistent with our model.

The orientation of the Moriya vector relative to the Mn-Mn bond is closely related to the twist and canting angles. While the discussion of these angles is rare in the literature \cite{Chizhikov12,Grytsiuk19,Hall21,Grytsiuk21} and the present study provides an experimental determination of their values, the orientation of ${\bf D}$ has been considered in a few theoretical works. There is no consensus.

A value $|\sigma| = 2.28$ has been estimated \cite{Dmitrienko12} based on a formula \cite{Keffer62} which forces ${\bf D}$ to be perpendicular to its bond, consistent with a first-principles calculation \cite{Katsnelson10}. The $|\sigma|$ value is about two orders of magnitude larger than measured. A more recent {\em ab initio\/} study of MnSi has estimated the ${\bf D}$ components \cite{Borisov21}.
The condition $|D^\alpha|/J \ll 1$ is satisfied, in agreement with our present results. However, with a calculated $D/J$ ratio of approximately 1/20, ${\bf D}$ is drawn roughly perpendicular to the bond. This appears inconsistent with our results which predict the Moriya vector to be almost parallel to the bond at minimum $D/J$ (see Fig.~\ref{DoJ_bond}). We should also note that (i) the  computed $J \approx 20$~meV value is approximately four times larger than found here \footnote{Note the difference between the convention of Ref.~\onlinecite{Borisov21} and ours: the factors 1/2 of Eq.~\ref{Hamiltonian} are absent in the quoted reference.}, (ii) the Mn moment is nearly three times larger than the accepted value, and (iii) the ratio $|D^\alpha|/J$ is sizably larger for next-nearest neighbors than for nearest neighbors. In fact, MnSi is a dual system as pointed out in the introduction. Here, we describe the interactions between the Mn moment localized components. We limit ourselves to nearest-neighbor interactions, so that we work with the  minimal Hamiltonian.

Remarkably, a microscopic, but not a first-principles, model has predicted the dominant contribution to ${\bf D}$ to be parallel to the corresponding bond, which is close to our experimental result when $D/J$ is minimum \cite{Hopkinson09}. Reference~\onlinecite{Hamann11} also considers ${\bf D}$ to be parallel to the bond. 

The ground state energy resulting from Eq.~\ref{Hamiltonian} does not depend on the ${\bf k}$ orientation, at least up to the second order expansion \cite{Chizhikov12,SM}. While ${\bf k} \parallel {\bf B}_{\rm ext}$ for $B_{\rm ext} \gtrsim 0.1$~T, it is parallel to $\langle 111\rangle$ in ZF \cite{Ishikawa76}. This suggests the  existence of an anisotropy term in the Hamiltonian. Two origins have been proposed: same-site energy and exchange. The former has been widely discussed \cite{Hopkinson09,Grigoriev15,Chizhikov21,Hall21}.  It  generates a spin gap \cite{Maleyev06}. Recalling the absence of such a gap at the sub-$10^{-7}$\,eV level \cite{Yaouanc05,Yaouanc20,Dalmas21}, the energy scale would be too small to be effective. 
Accounting for an exchange anisotropy would require at least one other exchange constant along with $J$, among a maximum of six \cite{Hall21}. A step forward in this direction could be a quantitative analysis of the phase diagram for which an anisotropy has been experimentally observed \cite{Ohkuma22}.

In conclusion, through a quantitative $\mu$SR spectra analysis, we have validated a nearest-neighbor spin  Hamiltonian picture made of the sum of isotropic Heisenberg and DM terms \cite{Chizhikov12,*[{A first approach to the determination of a spin Hamiltonian using $\mu$SR data has been presented in }] [{}] Maeter09}.Quantitative information has been derived for the Cartesian components of ${\bf D}$. An additional relation between these components is required for their full determination. Our analysis suggests the compound to be  still cubic in its magnetically ordered state, in contrast to a proposal \cite{Butenko10}.

Our microscopic magnetic picture could be of interest, for example, for an insight into the partial magnetic order observed above $T_{\rm c}$ under pressure \cite{Pfleiderer04}, or the mechanisms at play for the stabilization of the skyrmion crystal
\cite{Muhlbauer09a,Bogdanov89,Bogdanov94,Butenko10} and related topological Hall and Nernst effects \cite{Neubauer09,Oike22}. Understanding magnetic properties in terms of the crystal structure could be crucial for magnetic engineering. The microscopic mechanism leading to ${\bf k} \parallel \langle 111\rangle$ in ZF has still to be discovered. A local anisotropy can be excluded. We anticipate that the parameter $\sigma$, and therefore the ZF twist and canting, and ${\bf D}$,  could be slightly different from given here when ${\bf k} \parallel \langle 111\rangle$ in ZF is explained. 

We are grateful to M.\ Chshiev and A.\ Manchon for discussions about {\em ab initio\/} methods, S.\ Grytsiuk for drawing our attention to Ref.~\onlinecite{Borisov21}, and A.\ Maisuradze for a critical reading of the manuscript. Part of this work was performed at the GPS spectrometer of the Swiss Muon Source (Paul Scherrer Institut, Villigen, Switzerland).

\bibliography{reference,MnSi_Hamiltonian}

\end{document}